\documentclass[a4paper,prd,showpacs,aps,amsmath,amssymb]{revtex4}
\usepackage{graphicx}
\usepackage{graphics}
\usepackage{amsmath}
\usepackage{dcolumn}
\usepackage{amssymb}
\usepackage{bm}
\usepackage{footmisc}

\begin{document}
\title{Off-center coherent-state representation and an application to semiclassics}
\author{Fernando Parisio}
\affiliation{Instituto de F\'{\i}sica, Universidade Federal de
Alagoas, 57072-970, Macei\'o, Alagoas, Brazil and\\
Departamento de F\'{\i}sica, Universidade Federal de Pernambuco, 50670-901, Recife, Pernambuco, Brazil}

\begin{abstract}
By using the overcompleteness of coherent states we find an
alternative form of the unit operator for which the ket and the
bra appearing under the integration sign do not refer to the same
phase-space point. This defines a new quantum representation in
terms of Bargmann functions, whose basic features are presented. 
A continuous family
of secondary reproducing kernels for the Bargmann functions is
obtained, showing that this quantity is not necessarily unique for
representations based on overcomplete sets. We illustrate the
applicability of the presented results by deriving a semiclassical
expression for the Feynman propagator that generalizes the
well-known van Vleck formula and seems to point a way to cope with
long-standing problems in semiclassical propagation of localized
states.

\end{abstract}

\maketitle

\section{Introduction}
In addressing a quantum mechanical problem a crucial point is the choice of a convenient
representation, which is made operational by its closure relation.
These relations may be expressed
in a number of ways, e.g., in terms of position eigenstates $\{ |
x \rangle \}$, as an integration in configuration space. It is a
common belief that once a basis in the Hilbert space is chosen the
representation of the unit operator is unique. However, this is not
necessarily true if one deals with overcomplete basis, such as the
set of eigenstates $\{ |z \rangle \}$ of the annihilation
operator, the so-called canonical coherent states \cite{klauder1}.
Despite the fact that the label $z$ can take any complex value, a theorem
by Cahill \cite{cahill} asserts that the kets related to an arbitrary
convergent sequence $\{z_i\}$ of points in the $z$-complex plane
(${\cal C}$) suffice to generate the space of states. Therefore,
we have a large amount of redundancy in the full set and a general
state ket $|\psi \rangle$ may, in principle, be written in
different ways in terms of the vectors $\{|z \rangle \}$. An
immediate corollary is that any path in the complex plane enables
a complete quantum representation. Examples are the results given
in reference \cite{circle}, where only the vectors on a circle
($z=Re^{i \varphi}$) are used to express an arbitrary state; and the
line representation \cite{line}, for which only coherent states of
vanishing momentum are considered. Nevertheless, in these cases
the identity operator is not explicitly expressed. The redundant
nature of coherent states can also be made evident by an example
which is not encompassed by the theorem of Cahill, namely, the Wigner
lattice $z=\sqrt{\pi}(l+im)$ with $l$ and $m$ integers, that
constitutes a basis of the Hilbert space. Thus, in principle, the
unit operator does not have a unique representation in terms of
coherent states.

In this work we shall explore the overcompleteness from a different perspective, namely,
by focusing on the fact that the overlap between any two states $| z'
\rangle$ and $| z'' \rangle$ is non-vanishing and finite. In a more
technical terminology this means that the reproducing kernel of the
coherent-state representation is not a Dirac delta distribution.
As we
shall see, it is possible to use the full $z$-complex plane in a
non-standard way, where the operator under the integration sign in
the resolution of unity is
not the projector $| z \rangle \langle z|$.

In the next section we give an account of some basic properties of
coherent states. In section III we derive an alternative form of
the unit operator [see equation (\ref{unit1})], which, in turn,
gives rise to a broader way to deal with the analytic
representation associated to the coherent states. Basic properties
such as general state and operator representations, reproducing
kernels and constraints are presented and discussed.

This sort of
result may be
of use for, at least, two reasons. First, although the final result of a
calculation does not depend on which particular representation
we use, a convenient choice may lead to a drastic simplification in calculations.
Second, and most importantly, approximate results {\it do} depend on the
particular form of the identity operator used in the intermediate
steps. Perhaps the clearest example is the semiclassical
evaluation of quantum propagators via the stationary exponent
method. In section IV we use the new unit operator to
generalize the van Vleck semiclassical expression for the Feynman
propagator $\langle x''| \hat{K}(t)| x'\rangle$. In fact, we
obtain a family of semiclassical propagators parameterized by
a real number $\lambda$, of which the van Vleck formula is a
particular case ($\lambda=1$). While the original result involves
real classical paths that begin at $x'$ and end at $x''$, we show
that complex trajectories may appear in the semiclassical
evaluation of a position-position propagator and not only in
problems involving Gaussian initial states, as has been
reported so far \cite{adachi, klauder, parisio1, parisio2}.
Yet, within the realm of semiclassical approximations, since $\lambda$ is
a continuous parameter, it may be used in variational and optimization processes,
although we do not address these issues in the present work.
In the appendix we present alternative forms of the
unit operator whose application, however, must be made under
restricted conditions due to their weak convergence properties.


\section{The canonical coherent states}

The canonical coherent states are the eigenstates of the annihilation operator, $\hat{a}| z \rangle =z| z \rangle $.
All the unusual properties of this basis come from the non-hermiticity of $\hat{a}$, e. g., the spectral theorem
does not apply, neither the eigenvalues are real numbers nor are the eigenvectors mutually orthogonal.
In terms of harmonic oscillator eigenstates a normalized coherent state reads
\begin{equation}
\label{cs}
| z \rangle= e^{-\frac{1}{2}|z|^2}\;\sum_{n=0}^\infty \frac{z^n}{\sqrt{n!}}| n \rangle\;,
\end{equation}
with the complex label $z$ given by
\begin{equation}
z=\frac{q}{\sqrt{2}b}+i\frac{bp}{\sqrt{2}\hslash} \;,
\end{equation}
$q$ and $p$ being the expected values of position and momentum for the
state $| z \rangle$, and $\sqrt{2}b=\Delta q$ its position uncertainty.
The standard way to express the closure relation is in terms of a phase-space
integration:
\begin{equation}
\hat{I}=\int \frac{{\rm d}^2 z~}{\pi} \; |z \rangle \langle z|
\equiv \int\frac{{\rm d}q {\rm d}p}{2\pi\hbar}\;  |z \rangle \langle z|
\;.
\label{unit0}
\end{equation}
Since all states $| z \rangle$ are of minimum uncertainty their position
representation are given by wave functions that lead to Gaussian
probability densities,
\begin{equation}
\label{overlap}
\langle x| z\rangle=\frac{1}{\pi^{1/4}b^{1/2}}\; e^{-\frac{1}{2}(\frac{x}{b}-\sqrt{2}z)^2+\frac{1}{2}z(z-z^{*})}
=\frac{1}{\pi^{1/4}b^{1/2}}\; e^{ \frac{(x-q)^2}{2b^2}+\frac{i}{\hslash}p(x-\frac{q}{2})}\;,
\end{equation}
with a completely analogous relation holding for the momentum representation $\langle p_x| z\rangle$.

Finally, we briefly refer to the analytic representation
associated to the canonical coherent states, the so-called
Bargmann representation. In this formalism, a state ket
$|\psi\rangle$ is represented in phase space by its projection
onto a non-normalized coherent state $|z)=e^{\frac{1}{2}|z|^2}|z \rangle$,
in terms of which the standard resolution of unit is expressed as
\begin{equation}
\label{unitB0}
\hat{I}=
\int \frac{{\rm d}^2 z}{\pi} e^{-|z|^2} \; | z) (z|\;.
\end{equation}
The state of the system is completely determined by the entire
function $\psi(z^*)=( z|\psi \rangle$, i.e.,
\begin{equation}
\psi(z^*)=
\int \frac{{\rm d}^2 z'}{\pi} e^{-|z'|^2} \; ( z| z') (z'|\psi \rangle =\int \frac{{\rm d}^2 z'}{\pi} e^{-|z'|^2} \; ( z|z' ) \psi({z'}^*) \;.
\end{equation}
The reproducing kernel ${\cal K}(z^*,z')=( z|z' )= e^{z^*z'}$ plays the role of
the delta function in the position and momentum representations. Note,
however that the above relation represents an actual constraint to be
satisfied by the function $\psi(z^*)$, which has no parallel in the
other mentioned representations. We will describe further developments
on the Bargmann representation with the help of the results presented
in the next section.

\section{Off-center coherent-state representations}

The previous discussion motivates the inspection of the following kind of integral operator
\begin{equation}
\label{def}
\hat{A} =
\int \frac{{\rm d}^2 z}{\pi} \;\mu(z,z^*) \; | f(z,z^*)\rangle \langle z|\;,
\end{equation}
where the integration runs in the whole phase space (${\rm d}^2
z/\pi={\rm d}q{\rm d}p/2\pi \hslash$), $\mu$ is a weighting
function, and $f$ maps points on the complex plane ${\cal
C}$ to a subset ${\cal D}$ of ${\cal C}$. It is a key point to
realize that $| f(z,z^*)\rangle$ is itself a
coherent state with the same uncertainty in position as $|z
\rangle$, i. e., given an annihilation operator $\hat{a}$, if $\hat{a}| z \rangle =z| z \rangle $ then
$\hat{a}| f(z,z^*) \rangle =f(z,z^*)| f(z,z^*) \rangle $, where
$f$ can always be put in the form $f={\cal
Q}(q,p)/\sqrt{2}b+ib\,{\cal P}(q,p)/\sqrt{2}\hslash$. Note that,
for $f=z$ and $\mu=1$ we get the standard coherent-state
closure relation (\ref{unit0}).

Due to the mentioned non-vanishing overlap between distinct coherent states, we may hope that it is possible to obtain representations of $\hat{I}$ for which $f \neq z$ and, consequently, $\mu \ne 1$. In what follows we show that this is indeed the case by giving the explicit representation of a non-standard identity operator. Because of the non-equality between $f$ and $z$ we name this resolution of unity ``off-center''.
Notice that a definition analogous to equation (\ref{def}) involving position or momentum eigenstates, for example, could not result in the identity operator.

We recall that for vector spaces of bounded dimension it is a
necessary and sufficient condition for the equivalence $\hat{A}
\equiv \hat{I}$, that $\langle i | \hat{A} | j \rangle = A_{i,j} =
\delta_{i,j}$, for all pairs $i,j$, where $|i \rangle$ and $|j
\rangle$ belong to a discrete basis $\{|i \rangle \}$ of the
Hilbert space. Let $| \psi \rangle = \sum_i a_i |i \rangle$ be an
arbitrary ket, so $\hat{A}|\psi \rangle = \sum_i a_i \hat{A}|i
\rangle$. If $A_{i,j} = \delta_{i,j}$, then $\langle j
|\hat{A}|\psi \rangle= \sum_i a_i \langle j |\hat{A}|i \rangle
=a_j$. But $a_j=\langle j| \psi \rangle$, so $\hat{A}| \psi
\rangle$ and $| \psi \rangle$ have the same components in all
directions. Therefore $\hat{A}| \psi \rangle=| \psi \rangle$ for
an arbitrary ket $|\psi \rangle$, implying $\hat{A} \equiv
\hat{I}$. The converse is immediate. For infinite-dimensional
spaces $\langle i | \hat{A} | j \rangle = \delta_{i,j}$ is
certainly a necessary condition, but by no means a sufficient one.
In the appendix this will become particularly clear with two
explicit examples.

We focus our study on a very simple family of trial operators
\begin{equation}
\label{trial1}
\hat{A} =
\int \frac{{\rm d}^2 z}{\pi} \;\mu(z,z^*; \lambda) \; | \lambda z\rangle \langle z|\;,
\end{equation}
where $f=\lambda z$, with ${\cal D} = {\cal C}$, $\lambda$ being a real number.
Our task is to find out the weighting function $\mu$ that turns this expression into the unit operator.
Let $\{ | n \rangle \}$ be the harmonic oscillator basis associated to the annihilation operator $\hat{a}$ that defines
the set $\{ | z \rangle \}$. Note that if $| \lambda z \rangle$
is a coherent state associated to the {\it same} annihilation operator, its expected values of position and momentum are uniquely determined
by ${\cal Q}=\lambda q$ and  ${\cal P}=\lambda p$. Therefore, we have
\begin{equation}
\langle n |\hat{A}| m \rangle \equiv A_{n,m}=
\int \frac{{\rm d}^2 z}{\pi} \;\mu(z,z^*; \lambda) \; \langle n| \lambda z\rangle \langle z|m \rangle
=\frac{1}{\sqrt{n!m!}}\int \frac{{\rm d}^2 z}{\pi}\;\mu\; e^{-\frac{1}{2}|\lambda z|^2}\lambda^n z^n\; e^{-\frac{1}{2}|z|^2}{z^*}^m\;,
\end{equation}
where we used the relation (\ref{cs}). Writing the above expression in polar coordinates ($z=re^{i \varphi}$) we obtain
\begin{equation}
A_{n,m}=
\frac{1}{\sqrt{n!m!}}\int \frac{r{\rm d}\varphi {\rm d}r}{\pi} \;\mu \; \lambda^n \; e^{-\frac{1}{2}(\lambda^2+1)r^2}r^{n+m}\; e^{i\varphi(n-m)}
= \frac{2 \delta_{n,m}}{n!}\int_0^\infty {\rm d}r\;\mu\;r^{2n+1}  \lambda^n \; e^{-\frac{1}{2}(\lambda^2+1)r^2}\;.
\end{equation}
By writing $x=\lambda r^2$, we get
\begin{equation}
A_{n,m}
= \frac{ \delta_{n,m}}{\lambda n!}\int_0^\infty {\rm d}x\;\mu\;x^{n} \; e^{-\frac{1}{2}({\lambda}+\frac{1}{\lambda})x}\;,
\end{equation}
where the integral converges only if $\lambda>0$. Now, we see that the measure must be given by
\begin{equation}
\label{measure1}
\mu= \lambda \; e^{\frac{1}{2}({\lambda}+\frac{1}{\lambda})x}\; e^{-x}=\lambda \; e^{\frac{1}{2}(\lambda^2+1-2\lambda)r^2}\;,
\end{equation}
leading to $A_{n,m}=\delta_{n,m} \Gamma(n+1)/n!=\delta_{n,m}$.
Thus, we conclude that a good candidate to represent the resolution of unity is
\begin{equation}
\int \frac{{\rm d}^2 z}{\pi} \; \lambda \; e^{\frac{1}{2}(\lambda-1)^2|z|^2}  | \lambda z\rangle \langle z| \;,
\end{equation}
with $\lambda>0$, for it leads to the correct results whenever
convergence is guaranteed. Note that for $\lambda = 1$ we obtain
equation (\ref{unit0}). Now we have to show that the above
operator provides a consistent way to calculate the inner product
for all vectors in the Hilbert space, that is, we have to
demonstrate that
\begin{equation}
\label{auxiliary}
\int \frac{{\rm d}^2 z}{\pi} \; \lambda \; e^{\frac{1}{2}(\lambda-1)^2|z|^2} \; \langle \psi | \lambda z\rangle \langle z| \psi \rangle\;,
\end{equation}
converges to $\langle \psi | \psi \rangle$ for all vectors $| \psi \rangle$. Note that once we show that $|\langle \psi | \psi \rangle|<\infty$ we have $|\langle \psi | \phi \rangle|<\infty$ since $|\langle \psi | \phi \rangle|<|\langle \psi | \psi \rangle|$.
In order to give a rigorous proof we will follow arguments similar to the ones given in \cite{klauder2} to validate the standard coherent-state closure relation. Consider the following quantity
\begin{equation}
\label{finite}
\int_0^R \int_0^{2\pi} \frac{{\rm d}\varphi {\rm d}r}{\pi} \; \lambda \; e^{\frac{1}{2}(\lambda-1)^2r^2} \; \langle \psi | \lambda re^{i\varphi}\rangle \langle re^{i\varphi}| \psi \rangle\;.
\end{equation}
By using harmonic oscillator closure relations, and the fact that the above integration covers a finite region of phase space, justifying the interchange in the ordering of integrals and sums, one can write it as
\begin{eqnarray}
\nonumber
\sum_{n,m} \int_0^R \int_0^{2\pi} \frac{{\rm d}\varphi {\rm d}r}{\pi} \; e^{\frac{1}{2}(\lambda-1)^2r^2} \; \frac{\lambda^{n+1}r^{n+m+1}}{\sqrt{n!m!}}\; e^{-\frac{1}{2}(1+\lambda^2)|z|^2} \; e^{i\varphi(n-m)}\langle \psi | n \rangle \langle m| \psi \rangle \\
=\sum_{n} 2\int_0^R   \; \lambda^{n+1} \; e^{-\lambda r^2} \; \frac{r^{2n+1}}{n!} \; |\langle n| \psi \rangle|^2=\sum_n |\langle n| \psi \rangle|^2 \; \gamma_n\;,
\end{eqnarray}
where $\gamma_n$ can be rewritten as
\begin{equation}
\label{gamma}
\gamma_n=\frac{2}{n!}\int_0^{ R} {\rm d}r\; \lambda^{n+1} \; r^{2n+1}\;e^{-\lambda r^2}=\frac{1}{n!}\int_0^{\lambda R^2} {\rm d}x \; x^n\;e^{-x} \;,
\end{equation}
with $\lambda >0$. It is clear that for any positive $\lambda$ and real $R$ we have $0< \gamma_n <1$ and $\lim_{R \rightarrow \infty}\;
\gamma_n=1$. Then, we can write
\begin{equation}
\langle \psi | \psi \rangle = \sum_n |\langle n| \psi \rangle|^2 = \sum_n |\langle n| \psi \rangle|^2 \lim_{R \rightarrow \infty} \gamma_n= \lim_{R \rightarrow \infty}\sum_n |\langle n| \psi \rangle|^2 \; \gamma_n=\int \frac{{\rm d}^2 z}{\pi} \; \lambda \; e^{\frac{1}{2}(\lambda-1)^2|z|^2} \; \langle \psi | \lambda z\rangle \langle z| \psi \rangle\;,
\end{equation}
where we used the fact that $\sum_n |\langle n| \psi \rangle|^2 $
is absolutely convergent. This establishes the validity of
\begin{equation}
\label{unit1}
\hat{I}=\int \frac{{\rm d}^2 z}{\pi} \; \lambda \; e^{\frac{1}{2}(\lambda-1)^2|z|^2} \; | \lambda z\rangle \langle z|=\int \frac{{\rm d}^2 z}{\pi} \; \lambda \; e^{\frac{1}{2}(\lambda-1)^2|z|^2} \; |  z\rangle \langle \lambda z|
\end{equation}
as genuine representations of the unit operator in the whole
Hilbert space for $0< \lambda \le 1$. This limitation in the range
of $\lambda$ is due to the fact that definition (\ref{finite}) is
consistent for $\lambda \le 1$ only. For $\lambda > 1$ the complex
number $\lambda z$ associated to the ket in (\ref{finite}) will be
outside the integration region (whose boundary is $|z|=R$) for
$|z|>R/\lambda$. In order to guarantee that both, the labels of
the bra and the ket in integration (\ref{finite}) are inside the
integration region for all $r \le R $ we have to set $0< \lambda
\le 1$. Furthermore, despite $| \lambda z\rangle \langle z| \ne |
z\rangle \langle \lambda z|$, the second equality follows
immediately from the Hermiticity of the unit operator. Now, it is
quite simple to extend our result to $\lambda >1$. By taking
$w=\lambda z$ we get
\begin{equation}
\int \frac{{\rm d}^2 w}{\pi} \; \lambda^{-1} \; e^{\frac{1}{2}(\lambda^{-1}-1)^2|z|^2} \; |  w\rangle \langle \lambda^{-1} w|\;.
\end{equation}
Since $0< \lambda \le 1$, we have that $\pi^{-1}\int {\rm d}^2 w \; \sigma \; e^{\frac{1}{2}(\sigma-1)^2|z|^2} \; |  w\rangle \langle\sigma w|$
is valid for $1< \sigma \le \infty$ with $\sigma=\lambda^{-1}$. This is exactly relation (\ref{unit1}), which is, therefore, valid for any $\lambda >0$.

Finally, the inner product of two arbitrary vectors and the matrix elements of an operator $\hat{B}$ can be safely written as
\begin{equation}
\langle \phi | \psi \rangle=\int \frac{{\rm d}^2 z}{\pi} \; \lambda \; e^{\frac{1}{2}(\lambda-1)^2|z|^2} \; \langle \phi| \lambda z\rangle \langle z| \psi \rangle \;, \;\;\;\langle z |\hat{B}| z'\rangle =\int \frac{{\rm d}^2 z''}{\pi} \; \lambda \; e^{\frac{1}{2}(\lambda-1)^2|z''|^2} \;  \langle z|  \lambda z''\rangle \langle z''|\hat{B} | z' \rangle \;,
\end{equation}
respectively.
\subsection{Bargmann Representation: secondary reproducing kernels}
Let us analyze how the Bargmann functions are expressed via
relation (\ref{unit1}), which can also be written as
\begin{equation}
\label{unitB1}
\hat{I}=
\int \frac{{\rm d}^2 z}{\pi}\; \lambda \; e^{-\lambda|z|^2} \; | \lambda z) (z|\;.
\end{equation}
Then, we can write an arbitrary state as
\begin{equation}
\label{state2}
\psi(z^*)=\int \frac{{\rm d}^2 z'}{\pi} \; \lambda \; e^{-\lambda|z'|^2} \;  ( z | \lambda z') ( z'| \psi \rangle= \int \frac{{\rm d}^2 z'}{\pi} \; \lambda \; e^{-\lambda|z'|^2} \;{\cal K}(z^*,z';\lambda) \psi({z'}^*)\;,
\end{equation}
where the reproducing kernel of the representation is
\begin{equation}
{\cal K}(z^*,z';\lambda)= ( z | \lambda z')=e^{\lambda z^*z'}\; .
\end{equation}
By taking $| \psi \rangle = | \lambda z'' ) $ in equation (\ref{state2}) we show that ${\cal K}$
satisfies its own integral equation, i.e.,
\begin{equation}
{\cal K}(z^*,z'';\lambda)=\int \frac{{\rm d}^2 z'}{\pi} \; \lambda \; e^{-\lambda|z'|^2} \;{\cal K}(z^*,z';\lambda){\cal K}({z'}^*,z'';\lambda) \; .
\end{equation}
Once we identified the reproducing kernel by inspection of the last part of equation (\ref{state2}), we now show that ${\cal K}$ is not uniquely defined even in the case $\lambda=1$, in the sense that other phase-space functions also satisfy (\ref{state2}).

The fact that we are dealing with a continuous family of possible
representations allows us to write the following differential
relation for the kernel
\begin{equation}
\label{diff}
\frac{\partial^N {\cal K}}{\partial \lambda^N}=(z^*z')^N {\cal K} \;.
\end{equation}
Since the function $\psi(z^*)$ does not depend on $\lambda$, we can use relations (\ref{state2}) and (\ref{diff}) to write
\begin{equation}
\label{dif1}
\frac{{\rm d} \psi(z^*)}{{\rm d} \lambda}=\frac{\psi(z^*)}{\lambda}+\int \frac{{\rm d}^2 z'}{\pi} \; \lambda \; e^{-\lambda|z'|^2}
\;{\cal K}(z^*,z';\lambda)\;z'(z^*-{z'}^*)\;\psi({z'}^*)=0\;,
\end{equation}
which leads to
\begin{equation}
\label{int1}
\psi(z^*)=-\int \frac{{\rm d}^2 z'}{\pi} \; \lambda^2 \; e^{-\lambda|z'|^2}
\;{\cal K}(z^*,z';\lambda)\;z'(z^*-{z'}^*)\;\psi({z'}^*)\;.
\end{equation}
Differentiating (\ref{dif1}) once again with respect to $\lambda$ and using (\ref{int1}) we get
\begin{equation}
\label{int2}
\psi(z^*)=\int \frac{{\rm d}^2 z'}{\pi} \; \frac{\lambda^3}{2} \; e^{-\lambda|z'|^2}
\;{\cal K}(z^*,z';\lambda)\;[z'(z^*-{z'}^*)]^2\;\psi({z'}^*)\;.
\end{equation}
By induction we get the following general relation
\begin{equation}
\label{intN}
\psi(z^*)=\int \frac{{\rm d}^2 z'}{\pi} \; \lambda \;  e^{-\lambda|z'|^2}
{\rm K}^{(N)}(z^*,z';\lambda)\;\psi({z'}^*)\;,
\end{equation}
with
\begin{equation}
\label{kernel1a}
{\rm K}^{(N)}(z^*,z';\lambda)=\frac{\lambda^N}{N!}\;[z'({z'}^*-z^*)]^N\;{\cal K}(z^*,z';\lambda)\;,
\end{equation}
which defines a whole class of reproducing kernels,
with ${\rm K}^{(0)}={\cal K}$. Note, however, that the secondary kernels do not satisfy their own integral equations.
This is due to the fact that ${\rm K}^{(N)}$ can no longer be represented as an inner product for $N \ne 0$.
The above relation can be directly derived from the observation that
\begin{equation}
e^{-\lambda|z'|^2}\;[z'({z'}^*-z^*)]^N\;{\cal K}(z^*,z';\lambda)=[z'({z'}^*-z^*)]^N\;{\cal K}(z^*-{z'}^*,-z';\lambda)=(-1)^N\frac{\partial^N {\cal K}(z^*-{z'}^*,-z';\lambda)}{\partial \lambda^N}\;,
\end{equation}
where the last equality comes from relation (\ref{diff}). Therefore, we can re-write equation (\ref{intN}) as
\begin{equation}
\psi(z^*)=\frac{(-1)^N \lambda^{N+1}}{N!}\int \frac{{\rm d}^2 z'}{\pi} \; \frac{\partial^N {\cal K}(z^*-{z'}^*,-z';\lambda)}{\partial \lambda^N}\;\psi({z'}^*)=\frac{(-1)^N \lambda^{N+1}}{N!}\frac{\partial^N }{\partial \lambda^N}\int \frac{{\rm d}^2 z'}{\pi} \;e^{-\lambda|z'|^2}\;
{\cal K}(z^*,z';\lambda) \;\psi({z'}^*)\;.
\end{equation}
By the primary relation (\ref{state2}) we see that the last integral is simply $\psi(z^*)/\lambda$, which finishes the proof, since $(-1)^N \lambda^{N+1}\;\partial^N (\lambda)^{-1}/\partial \lambda^N=N!$ .
As a matter of fact, we note that the first equality in the above relation defines a family of reproducing kernels for the uniform
measure:
\begin{equation}
\label{kernel1b}
{\rm \tilde{K}}^{(N)}(z^*-{z'}^*,-z';\lambda)=(-1)^N\frac{\lambda^{N+1}}{N!}\;\frac{\partial^N {\cal K}(z^*-{z'}^*,-z';\lambda)}{\partial \lambda^N}\;,
\end{equation}
for any positive $\lambda$ and $N=0,1,2$...
This makes clear that the infinity of ways to represent a reproducing integral equation for a general Bargmann state has its roots
in the overcompleteness of coherent states, and not in a non-trivial measure.

\section{Generalization of the van-Vleck formula}
As an initial application of equation (\ref{unit1})
we show that semiclassical results may be quite sensitive to the particular kind of identity operator
one employs in the intermediate steps of asymptotic calculations. We start by recalling the van Vleck semiclassical formula for the
Feynman propagator $\langle x''| \hat{K}(t)| x'\rangle$,
\begin{equation}
\label{vv}
\langle x''| \hat{K}(t)| x'\rangle_{VV}=\frac{e^{\frac{i}{\hslash}S}}{b \sqrt{2\pi i m_{qp}}} \;,
\end{equation}
where $S=S(x',x'';t)$ is the action integral of the classical trajectory that starts at $x'$ and ends at $x''$ in a time interval $t$, and $m_{qp}$ is an element of the stability matrix. This matrix specifies how a small rectangular spot in phase space with sides $\delta q_0$ and $\delta p_0$ develops in time, in the linear regime. Explicitly we have
\begin{equation}
\delta q_t= m_{qq}\; \delta q_0 +\frac{b^2}{\hslash} m_{qp} \;\delta p_0 \; , \;\; \delta p_t= \frac{\hslash}{b^2}m_{pq}\; \delta q_0 + m_{pp}\; \delta p_0 \; .
\end{equation}
In analogy to the exact quantum mechanical relation
\begin{equation}
\label{exact}
\langle x''| \hat{K}(t)| x'\rangle=\langle x''| \hat{K}(t) \hat{I}| x'\rangle=\int \frac{{\rm d}^2 z}{\pi} \;  \langle x''|\hat{K}(t)|z \rangle \;\langle z|x' \rangle \;,
\end{equation}
it has been demonstrated that the following semiclassical relation holds \cite{marcus}
\begin{eqnarray}
\langle x''| \hat{K}(t)| x'\rangle_{VV}=\int \frac{{\rm d}^2 z}{\pi} \;  \langle x''|\hat{K}(t)|z \rangle_{Heller} \;\langle z|x' \rangle\;,
\end{eqnarray}
where we used relation (\ref{overlap})
and the Heller thawed approximation that is given by
\begin{equation}
\label{heller}
\langle x''|\hat{K}(t)|z \rangle_{Heller}=\frac{\pi^{-1/4} b^{-1/2}}{\sqrt{ m_{qq}+im_{qp}}}\;e^{-\frac{\xi}{2b^2}(x''-q_t)^2}
e^{\frac{i}{\hslash}\left[S+p_t(x''-q_t)+\frac{1}{2}qp\right]} \;,
\end{equation}
with $q_t$ and $p_t$ being the final points of the real trajectory that begins at $q$ with momentum $p$, and $\xi=(m_{pp}-im_{pq})/(m_{qq}+im_{qp})$.

Notice that equation (\ref{exact}) was obtained with the use of the standard closure relation (\ref{unit0}).
In complete analogy to
this procedure, if $\hat{A}$ is the identity operator, one can use relation (\ref{def}) to write
\begin{eqnarray}
\label{exact2}
\nonumber
\langle x''| \hat{K}(t)| x'\rangle=\int \frac{{\rm d}^2 z}{\pi} \;\mu(z,z^*)  \langle x''|\hat{K}(t)|f(z) \rangle \;\langle z|x' \rangle \\
=\int \frac{{\rm d}^2 z}{\pi} \;[\mu(z,z^*)]^*  \langle x''|\hat{K}(t)|z \rangle \;\langle f(z)|x' \rangle\;
\end{eqnarray}
and define an alternative family of semiclassical propagators from
relation (\ref{unit1}),
\begin{equation}
\langle x''| \hat{K}(t)| x'\rangle_{sc}=\int \frac{{\rm d}^2 z}{\pi} \; \lambda\; e^{\frac{1}{2}(\lambda-1)^2|z|^2}
\langle x''|\hat{K}(t)|z \rangle_{Heller} \;\langle \lambda z|x' \rangle\;,
\end{equation}
where we used the second line of Eq. (\ref{exact2}), which is
valid due to the hermiticity of $\hat{I}$. The integral to be
calculated can be written as
\begin{equation}
\langle x''| \hat{K}(t)| x'\rangle_{sc}=\frac{\lambda}{b \sqrt{\pi}}\int \frac{{\rm d} q {\rm d} p}{2\pi \hslash} \;
\frac{e^{\Gamma(q,p,\lambda)}}{\sqrt{ m_{qq}+im_{qp}}}
 \;,
\end{equation}
where
\begin{eqnarray}
\nonumber
\Gamma(q,p,\lambda)= -\frac{\xi}{2b^2}(x''-q_t)^2
+\frac{i}{\hslash}\left[S+p_t(x''-q_t)+\frac{1}{2}qp\right]\\
-\frac{1}{2}({\lambda}^2+2\lambda-1)\frac{q^2}{2b^2}+\frac{1}{2}(\lambda-1)^2\frac{b^2p^2}{2 \hslash^2}
+\frac{i {\lambda}^2}{2 \hslash}qp+\frac{\lambda x'}{b^2}q-\frac{i \lambda x' }{\hslash}p-\frac{{x'}^2}{2 b^2}\;.
\end{eqnarray}
Now we evaluate the phase-space integral via the saddle point method. The stationary conditions $\Gamma_q=\partial \Gamma/ \partial q=0$ and
$\Gamma_p=\partial \Gamma/ \partial p=0$, after some algebra, lead to
\begin{equation}
(1-{\lambda}^2)\frac{v_0}{\sqrt{2}}+\lambda\frac{(x'-q_0)}{b}+\frac{1}{m_{qq}+im_{qp}}\frac{(x''-q_t)}{b}=0\;,
\end{equation}
and
\begin{equation}
(1-{\lambda})^2\frac{v_0}{\sqrt{2}}-\lambda\frac{(x'-q_0)}{b}+\frac{1}{m_{qq}+im_{qp}}\frac{(x''-q_t)}{b}=0\;,
\end{equation}
where we have to consider the action $S=S(q_0,q_t(q_0,p_0);t)$ as an
implicit function of $q_0$ and $p_0$ and use the relations
$\partial S/\partial q_0 = -p_0$ and $\partial S/\partial q_t =
p_t$. The subscript ``0'' denotes the stationary point in phase
space and $v_0\equiv q_0/\sqrt{2}b-ib\,p_0/\sqrt{2}\hslash$. It must be
realized that to obtain the above results, and in the rest of the
calculations, we do not take into account variations of
$m_{qq}$, $m_{qp}$, and $\xi$ because they already involve second
derivatives of the action integral, e. g., $m_{qp} \propto (\partial^2 S/\partial q_0 \partial q_t)^{-1}$.
Otherwise one would get terms
with order higher than second (for a detailed discussion of the
saddle point method applied to the evaluation of
semiclassical propagators see ref. \cite{marcus}). The previous
conditions can be rewritten as
\begin{equation}
\label{cc}
({\lambda}-1)v_0=\frac{\sqrt{2}}{b}(x'-q_0)\;,\;\; \mbox{and}\;\;({\lambda}-1)v_0=\frac{\sqrt{2}}{m_{qq}+im_{qp}}\frac{(x''-q_t)}{b}\;,
\end{equation}
which make clear that only for $\lambda=1$ we have $q_0=x'$ and $q_t=x''$, the van Vleck boundary conditions. For other values of $\lambda$ the classical path that comes from the stationary condition must have, both, complex position and momentum. Note that, in this case, $v_0$ does not coincide with $z_0^*=q_0^*/\sqrt{2}b-ib\,p_0^*/\sqrt{2}\hslash$. Complex trajectories naturally arise in the semiclassical evaluation of propagators involving coherent states, e. g. $\langle x| \hat{K}(t)| z \rangle$ \cite{adachi, klauder, parisio1, parisio2}, since a trajectory that begins with momentum $p$ and position $q$ and ends at $x$ is clearly over-specified.  We now see that they can also appear in the simpler case of a position-position propagator.

The next step is to expand the exponent around $q_0$ and $p_0$ up to second order. We write
\begin{equation}
\Gamma= \Gamma_0 +\frac{b^2}{2}\Gamma_{qq}\tilde{Q}^2+\hslash \Gamma_{qp}\tilde{Q} \tilde{P} +\frac{\hslash^2}{2b^2}\Gamma_{pp}\tilde{P}^2\;,
\end{equation}
where $\tilde{Q}=(q-q_0)/b$ and $\tilde{P}=b(p-p_0)/\hslash$, and $\Gamma_0$ is the exponent evaluated at the stationary point. The semiclassical propagator becomes
\begin{equation}
\label{gaussint}
\langle x''| \hat{K}(t)| x'\rangle_{sc}=\frac{\lambda}{b \sqrt{\pi}}\frac{e^{\Gamma_0}}{\sqrt{m_{qq}+im_{qp}}}
\int \frac{{\rm d} \tilde{Q} {\rm d} \tilde{P}}{2\pi} \; e^{\frac{b^2}{2}\Gamma_{qq}\tilde{Q}^2+\hslash \Gamma_{qp}\tilde{Q} \tilde{P} +\frac{\hslash^2}{2b^2}\Gamma_{pp}\tilde{P}^2}
 \;,
\end{equation}
where the second derivatives of $\Gamma$ read
\begin{equation}
\Gamma_{qq}=\frac{1}{2b^2}(1-2\lambda-{\lambda}^2)-\frac{1}{b^2}\frac{m_{qq}}{m_{qq}+im_{qp}}\;,
\end{equation}
\begin{equation}
\Gamma_{qp}=\Gamma_{pq}=\frac{i}{2\hslash}({\lambda}^2-1)-\frac{1}{\hslash}\frac{m_{qp}}{m_{qq}+im_{qp}}\;,
\end{equation}
and
\begin{equation}
\Gamma_{pp}=\frac{b^2}{2\hslash^2}({\lambda}-1)^2-\frac{i b^2}{\hslash^2}\frac{m_{qp}}{m_{qq}+im_{qp}}\;.
\end{equation}
The integral (\ref{gaussint}) is convergent if the matrix associated to the quadratic form in the exponent has both eigenvalues
($\sigma_{\pm}$) with a negative real part. After some algebra we find $(\sigma_{\pm})=\lambda[-1\pm (m_{qq}-im_{qp}/\sqrt{m_{qq}^2+m_{qp}^2}]$,
that satisfy the convergence condition.
We get
\begin{equation}
\langle x''| \hat{K}(t)| x'\rangle_{sc}=\frac{\lambda}{b \sqrt{\pi}}\frac{e^{\Gamma_0}}{\sqrt{m_{qq}+im_{qp}}} \;
\frac{1}{\sqrt{\hslash^2(\Gamma_{qq}\Gamma_{pp}-\Gamma_{qp}^2)}}\;.
\end{equation}
After some simple manipulations, where we used $\det[m]=1$, one obtains
\begin{equation}
\langle x''| \hat{K}(t)| x'\rangle_{sc}=\frac{e^{\Gamma_0}}{b \sqrt{2\pi i m_{qp}}}\;,
\end{equation}
which presents a pre-factor that is formally identical to that of
the van Vleck expression (\ref{vv}). The argument of the
exponential, however, is quite distinct. With the help of the
boundary conditions (\ref{cc}) we get the expression
\begin{equation}
\label{gamma0}
\Gamma_0= -\frac{1}{2b^2}({x'}^2-q_0^2)+ \frac{1}{b^2}\left(\frac{\lambda+1}{\lambda-1}\right)(x'-q_0)^2
-\frac{\xi}{2b^2}(x''-q_t)^2+\frac{i}{\hslash}[S+p_t(x''-q_t)]\;.
\end{equation}
The final result is then
\begin{equation}
\label{finalprop}
\langle x''| \hat{K}(t)| x'\rangle_{sc}=\frac{e^{\frac{i}{\hslash}S}}{b \sqrt{2\pi i m_{qp}}}\;
e^{-\frac{1}{2b^2}({x'}^2-q_0^2)+ \frac{1}{b^2}\left(\frac{\lambda+1}{\lambda-1}\right)(x'-q_0)^2
-\frac{\xi}{2b^2}(x''-q_t)^2+\frac{i}{\hslash}p_t(x''-q_t)}\;.
\end{equation}
In general, the application of the above expression to a
particular system provides different results for distinct values
of $\lambda$, all semiclassically valid. There are, however, two
exceptions: the free particle and the harmonic oscillator, for
which all consistent second order semiclassical expressions must
coincide with the exact quantum result. In the next subsection we
show this explicitly for the free particle.

It is known that, for
a general anharmonic system, the van Vleck formula presents
spurious divergences when the classical path passes through a
caustic ($m_{qp}=0$). Since in our general expression the classical trajectory itself
depends on $\lambda$, the position and time at which a caustic
occurs are also $\lambda$-dependent. This suggests that one can
construct well behaved semiclassical propagator by
combining two or more expressions given by the family (\ref{finalprop}) with
different $\lambda$'s, conveniently chosen to avoid caustics. The
price to be paid is to deal with complex paths and connection conditions.
Note that for
$\lambda = 1$ the first part of equation (\ref{finalprop}) is exactly
the van Vleck propagator which is calculated for the real
trajectory with $q_0=x'$ and $q_t=x''$, while the second part is
equals to $1$, because the exponent goes to zero. This is also
valid for the term containing
$(\lambda-1)^{-1}$ since, according to Eq. (\ref{cc})
$(x'-q_0)^2 \propto (\lambda-1)^2$. Therefore we always have
$\langle x''| \hat{K}(t)| x'\rangle_{sc} = \langle x''|
\hat{K}(t)| x'\rangle_{VV}$, for $\lambda=1$.

A clarification is now in order. It might look awkward to have a
semiclassical propagator which depends on a parameter that is
neither present in the Hamiltonian of the system nor depends on
$\hslash$. However, we must keep in mind that $\lambda$ is not a
physical parameter but, rather, a mathematical one which is
present because of the redundancies associated to the
coherent-state representation. Our result gives an explicit
example of the fact that, very often, there is an infinity of
semiclassical formulas corresponding to the same quantity in
quantum mechanics. An already classical example is the two forms
of the coherent-state path integral given by Klauder and
Skagerstam \cite{klauder1}. In the first form the Hamiltonian that
determines the paths is $H(z^*,z)=\langle z | \hat{H}| z \rangle$,
while in the second form it is described by $h(z^*,z)$, with
\begin{equation}
\hat{H}=\int \frac{{\rm d}^2 z}{\pi} \; h(z^*,z) \; | z\rangle \langle z|\;.
\end{equation}
The point is that we find $H \ne h$ \cite{marcus}, although
they coincide in the classical limit, i.e., $H=H_c+{\cal
O}(\hslash)$ and $h=H_c+{\cal O}(\hslash)$, where $H_c$ is the
classical Hamiltonian. Besides, the function $h$ is not unique \cite{klauder1}.
Another example is given in
\cite{parisio1}, where the underlying classical dynamics from
which the semiclassical coherent-state propagators are derived is
governed by an effective Hamiltonian, whose smoothing parameter is
arbitrary and may vary continuously. Therefore, the Hamiltonian
operator (even with a specific ordering) present in a quantum
propagator does not uniquely determines its semiclassical
counterpart.

Before closing this work, we illustrate the use of formula (\ref{finalprop}) along with the boundary conditions (\ref{cc}) in the simple case of
a free particle with mass $m$. This also serves as a test of consistency for our result, since, for this system, we must obtain the exact quantum propagator for any $\lambda$.
We start with the solutions of the equations of motion $q_t=q_0+p_0 t/m$ and $p_t=p_0$,
which lead to the following elements for the stability matrix $m_{qq}=m_{pp}=1$, $m_{qp}=\hslash t/m b^2$, and $m_{pq}=0$.
The action integral is simply given by
\begin{equation}
S=\int^f_i(p{\rm d}q-H{\rm d}t)=\frac{p_0^2}{2m}t\;,
\end{equation}
where $H$ is the Hamiltonian function. The solutions of the equations of motion along with equations (\ref{cc}) enable us to write $q_t$, $q_0$, and $p_0$ in terms of $x'$, $x''$, and $t$.
The results are
\begin{equation}
q_0=\frac{1}{2 \lambda}\left[ (\lambda+1)x'+i(\lambda-1)\frac{mb^2}{\hslash t}(x''-x')\right]\;,
\end{equation}
\begin{equation}
q_t=\frac{1}{2 \lambda}\left\{ (\lambda+1)x''+i(\lambda-1)\left[\frac{mb^2}{\hslash t}(x''-x')-\frac{\hslash t}{m b^2}x'\right]\right\}\;,
\end{equation}
and
\begin{equation}
p_t=p_0=\frac{m}{2 \lambda t}\left[ (\lambda+1)(x''-x')-i(\lambda-1)\frac{\hslash t}{mb^2}x'\right]\;.
\end{equation}
Note that we have $q_0=x'$, $q_t=x''$, and $p_t=p_0=m(x''-x')/t$, for $\lambda=1$. Substituting all these quantities in the general expression for
$\Gamma_0$, equation (\ref{gamma0}), we simply get $\Gamma_0=im (x''-x')^2/2 \hslash t$. The final result for the semiclassical propagator of the
free particle is independent of $\lambda$ and reads
\begin{equation}
\langle x''| \hat{K}(t)| x'\rangle_{sc}=\sqrt{\frac{m}{2 \pi i \hslash t}}e^{\frac{im }{2 \hslash t}(x''-x')^2}\;,
\end{equation}
that coincides with the exact quantum result, as expected. Of course, the use of $\lambda \ne 1$ in this very simple case
only brings extra complications. The point is that, besides the caustic problem, in situations where it is not easy or even possible to determine the classical paths that satisfy
the van Vleck boundary conditions, we expect that an appropriate choice of $\lambda$ may simplify the problem.
\section{Conclusions and perspectives}
In this work we presented an off-center coherent-state
identity operator, showing that even when the whole phase space
is considered, there is no unique representation for the closure
relation in terms of the set $\{|z \rangle \}$. This property has enabled
the development of an alternative way to express the mathematical quantities
involved in the Bargmann representation as well as the derivation of extra
conditions that the function $\psi(z^*)$ has to meet. As a consequence we found
a family of reproducing kernels indexed by one continuous parameter $\lambda$ and an
integer $N$.
The potential
applicability of relation (\ref{unit1}) was illustrated for a
simple example in the field of semiclassics, which may enable the
construction of well behaved semiclassical position-position
propagators. We note that similar arguments can be used to derive
yet another form for the semiclassical Feynman propagator
starting from relation (\ref{unit1}) and the first line of relation (\ref{exact2}).
In fact, the sort of
procedure we used in section III is far from being exhausted. The
same technique can be used to generalize semiclassical
approximations involving Gaussian states, e.g., the propagator
$\langle x | \hat{K}(t)| z \rangle$, which is of
importance in different fields \cite{marcus,gau1,gau2,gau3}.
For this class of propagators the outcome of the
stationary exponent method, without any further approximation,
always involves complex trajectories and the root search problem 
\cite{marcus,novaes1,novaes2}.
In this context it would be a major simplification if one could 
adjust the parameter $\lambda$ in order to get an
initial value representation (IVR), or at least, to simplify the root
search problem.
Finally, it is also possible to use expression
(\ref{unit1}) in the direct evaluation of
propagators via path integrals, in the spirit of references
\cite{parisio1,marcus}. Some of these topics are presently under
investigation.

\acknowledgments
The author thanks M. A. M. de Aguiar, F. Brito, and M. Novaes for their suggestions on this manuscript.
This work was partially supported by the Brazilian agencies CNPq and FACEPE (APQ 0800-1.05/06).
\section{Appendix: alternative identities with conditional convergence}
In this appendix we analyze the extension of equation
(\ref{unit1}) when $\lambda$ is allowed to be complex and a
further example of ``pathological'' representation of the unit
operator. We stress that in both cases the resulting operators are
not unconditionally convergent. If we use a complex $\lambda$ in
the arguments presented in section II, the integration path in
(\ref{gamma}) is no longer on the real axis and the extension of
the initial variable $r$ into the complex plane would have to be
justified. The possible resolution of unit is
\begin{equation}
\label{unit1b}
\int \frac{{\rm d}^2 z}{\pi} \; \lambda \; e^{\frac{1}{2}(|\lambda|^2+1-2\lambda)|z|^2} \; | \lambda z\rangle \langle z|\;,
\end{equation}
with ${\rm Re}(\lambda)>0$, ${\cal Q}=(\lambda +
\lambda^{*})q/2+i(\lambda - \lambda^{*})b^2p/\hslash$, and  ${\cal
P}=(\lambda + \lambda^{*})p/2-i(\lambda - \lambda^{*})\hslash
q/b^2$. The important point is that in analogy to equation
(\ref{gamma}) we have $n!\gamma_n= \gamma(n+1,\lambda R^2)$, where
$\gamma(a,b)$ is the incomplete Gamma function. However, with the
complex argument $\lambda R^2$, for a given $n$, we may have
values of $R$ for which ${\rm Re}(\gamma_n)>1$. For definiteness
take $n=0$ and $\lambda=e^{i \pi/4}$, by setting $R^2=\sqrt{2}\pi$
we get ${\rm Re}(\gamma_n)=1+e^{-\pi} \approx 1,04$. For the
arguments presented in the body of the paper (in the case of real
$\lambda$) to be valid here we should have necessarily $0<{\rm
Re}(\gamma_n) \le 1$. Thus, unconditional convergence is not
guaranteed. As an example of successful application of
(\ref{unit1b}), let us consider the normalization of an arbitrary
{\it squeezed} state $| w \rangle$, with
$w=q'/\sqrt{2}B+iB\,p'/\sqrt{2}\hslash$. It is easy to show that
\cite{parisio1}
\begin{equation}
\langle w| z \rangle= \sqrt{\frac{2bB}{b^2+B^2}}\;e^{-\frac{1}{2}|w|^2-\frac{1}{2}|z|^2+\frac{1}{2}\left( \frac{B^2-b^2}{b^2+B^2}\right)(z^2-{w^*}^2)+\left( \frac{2bB}{b^2+B^2}\right)w^*z}\;.
\end{equation}
Therefore, if one uses identity (\ref{unit1b}) the inner product becomes
\begin{equation}
\langle w| w \rangle=\int \frac{{\rm d}^2 z}{\pi} \lambda e^{\frac{1}{2}(|\lambda|^2+1-2\lambda)|z|^2} \; \langle w| \lambda z\rangle \langle z|w \rangle= \frac{2 \lambda bB}{b^2+B^2}\; e^{-|w|^2-\frac{1}{2}\left( \frac{B^2-b^2}{b^2+B^2}\right)(w^2+{w^*}^2)}\;J\;,
\end{equation}
where
\begin{equation}
J=\int \frac{{\rm d}^2 z}{\pi} \; e^{-\lambda|z|^2+\frac{\lambda^2}{2}\left( \frac{B^2-b^2}{b^2+B^2}\right)z^2+\frac{1}{2}\left( \frac{B^2-b^2}{b^2+B^2}\right){z^*}^2+\left( \frac{2\lambda bB}{b^2+B^2}\right)w^*z+\left( \frac{2bB}{b^2+B^2}\right)wz^*}= \frac{b^2+B^2}{2 \lambda bB}\; e^{|w|^2+\frac{1}{2}\left( \frac{B^2-b^2}{b^2+B^2}\right)(w^2+{w^*}^2)} \;,
\end{equation}
whose convergence is guaranteed if $2 \lambda b^2/(b^2+B^2)$ and
$2 \lambda B^2/(b^2+B^2)$ have both positive real parts, which is
equivalent to ${\rm Re}(\lambda)>0$. This
leads to the correct normalization $\langle w| w \rangle=1$ for all values of the squeezing parameter $B$.
Nevertheless, expression (\ref{unit1b}) is not convergent, e. g., for matrix elements of
non-normalizable kets like $\langle x'| x \rangle$. In this case we should satisfy the extra condition
${\rm Re}(\lambda)>{\rm Im}(\lambda)$, which shows that the convergence is, in general, only conditional.

Now let us address the operator involving a ket that is restricted to the unit circle while the bra runs over the entire
phase space:
\begin{equation}
\label{trial2}
\hat{A} =
\int \frac{{\rm d}^2 z}{\pi} \;\mu(z,z^*) \; | z/|z|\rangle \langle z|\;.
\end{equation}
We have $f=e^{i{\rm Arg}(z)}=e^{i \varphi}$ with ${\cal D}$
being the unit circle. Accordingly, ${\cal Q}=q\,
(q^2/2b^2+b^2p^2/2\hslash^2)^{-1/2}$ and ${\cal
P}=p\,(q^2/2b^2+b^2p^2/2\hslash^2)^{-1/2}$. We will use once more
the harmonic oscillator basis in order to determine the
appropriate weighting function. We have
\begin{equation}
A_{n,m}=
\int \frac{{\rm d}^2 z}{\pi} \;\mu(z,z^*) \; \langle n| z/|z| \rangle \langle z|m \rangle
=\frac{1}{\sqrt{n!m!}}\int \frac{{\rm d}^2 z}{\pi}\; \mu \; e^{-\frac{1}{2}\left|\frac{z}{|z|}\right|^2} \left(\frac{z}{|z|}\right)^n\; e^{-\frac{1}{2}|z|^2}{z^*}^m\;.
\end{equation}
Passing again to polar coordinates one gets
\begin{equation}
A_{n,m}=
\frac{1}{\sqrt{n!m!}}\int \frac{r{\rm d}\varphi {\rm d}r}{\pi} \; \mu \;  \; e^{-\frac{1}{2}(r^2+1)}r^{m}\; e^{i\varphi(n-m)}
= \frac{2 \delta_{n,m}}{\sqrt{n!m!}}\int_0^\infty {\rm d}r \; \mu \; r^{m+1}\; e^{-\frac{1}{2}(r^2+1)}\;.
\end{equation}
By setting $\mu= \frac{1}{2r} \; e^{\frac{1}{2}(r-1)^2}$,
we obtain $A_{n,m}=\delta_{n,m}$.
The possible closure relation is then
\begin{equation}
\label{unit2}
\int \frac{{\rm d}^2 z}{\pi} \; \frac{1}{2|z|} \; e^{\frac{1}{2}(|z|-1)^2}\;| z/|z|\rangle \langle z|\;.
\end{equation}
Note that there is no actual singularity since ${\rm
d}^2z/|z|={\rm d} \varphi {\rm d}r$ as it is clear from the Bargmann
representation
\begin{equation}
\int \frac{{\rm d}^2 z}{\pi} \frac{1}{2|z|} \; e^{-|z|}\; | z/|z|) (z|=
\int \frac{{\rm d} \varphi{\rm d}r}{2\pi} e^{-r} \; |e^{i \varphi}) (re^{i \varphi}|\;.
\end{equation}
In showing the unconditional convergence of (\ref{unit1}) we defined a quantity, see equation (\ref{auxiliary}),
that would correspond in the present case to
\begin{equation}
\int _0^R
\int_0^{2\pi} \frac{{\rm d} \varphi{\rm d}r}{2\pi} e^{-r} \; \langle \psi|e^{i \varphi}) (re^{i \varphi}| \psi \rangle\;,
\end{equation}
which is meaningless in the region $r<1$, because $|e^{i
\varphi})$ always has $r=1$. Again, the arguments given in the
body of the paper cannot be repeated. In fact, it is quite easy to
find well behaved quantities that present spurious divergence when
(\ref{unit2}) is employed. The reader can easily show that
application of (\ref{unit2}) to the evaluation of $\langle w | w
\rangle$ yields convergent results only if $B>b$. Thus, expression
(\ref{unit2}) has very limited convergence properties and must be
seen as a formal result that requires a careful analysis to be
used in practice. The convergence of $\langle \psi|\hat{A}| \psi
\rangle $ can be guaranteed only if $| \psi \rangle$ is given
either by a finite superposition of number states or coherent
states.

We remark that (\ref{unit2}) is not directly related to the circle representation given in \cite{circle}. In the present case the integration still runs over the whole phase space, only the ket $| z/|z| \rangle=| e^{i\varphi} \rangle$ being restricted to the unit circle. Nevertheless, the circle representation can be obtained from our result. More specifically, using relation (\ref{unit2}) we can write a Fock state as
\begin{equation}
| n \rangle= \int \frac{{\rm d}^2 z}{\pi} \; \frac{1}{2|z|} \; e^{\frac{1}{2}(|z|-1)^2}\;| z/|z|\rangle \langle z| n \rangle=
\frac{e^{\frac{1}{2}}}{2\pi\sqrt{n!}}\int_0^{2 \pi} {\rm d} \varphi \left[\int_0^{\infty} {\rm d} r\; r^n\; e^{-r} \right]\;e^{-in \varphi}|e^{i \varphi}\rangle
\;,
\end{equation}
So,
\begin{equation}
| n \rangle=\frac{\sqrt{n!\,e}}{2\pi}\int_0^{2 \pi}{\rm d} \varphi\;e^{-in \varphi}|e^{i \varphi}\rangle\;,
\end{equation}
which is the unit circle representation of a number state \cite{circle,circle2}.


\begin{thebibliography}{99}

\bibitem{klauder1} J. R. Klauder and B. Skagerstam,
{\it Generalized Coherent Sattes and their Applications}, World Scientific-Singapore (1985).

\bibitem{cahill} K. E. Cahill, Phys. Rev. {\bf 138}, B1566 (1965).

\bibitem{circle} P. Domokos, P. Adam, and J. Janszky, Phys. Rev. A {\bf 50},
4293 (1994).

\bibitem{line} J. Janszky and An. V. Vinogradov, Phys. Rev. Lett. {\bf 64}, 2771 (1990).

\bibitem{adachi} S. Adachi, Ann. Phys. {\bf 195}, 45 (1989).

\bibitem{klauder} J. R. Klauder, Phys. Rev. D {\bf 19}, 2349 (1979).

\bibitem{parisio1} F. Parisio and M. A. M. de Aguiar, Phys. Rev. A {\bf 68}, 062112 (2003).

\bibitem{parisio2} F. Parisio and M. A. M. de Aguiar, J. Phys. A: Math. Gen. {\bf 38}, 9317 (2005).

\bibitem{klauder2} J. R. Klauder and E. C. G. Sudarshan,
{\it Quantum Optics}, W. A. Benjamin-New York (1968).

\bibitem{bargmann} V. Bargmann  Comm. on Pure and Appl. Math. {\bf 14}, 187 (1961).

\bibitem{marcus} M. Barager, M. A. M. de Aguiar, F. Keck, H. J. Korsch, and B. Schellhaass, J. Phys. A: Math. Gen. {\bf 34}, 7227 (2001).

\bibitem{circle2} J. Janszky, P. Domokos, and P. Adam, Phys. Rev. A {\bf 48},
2213 (1993).

\bibitem{gau1} R. N. Maia, F. Nicacio, R. O. Vallejos, and F. Toscano, Phys. Rev. Lett. {\bf 100},
184102 (2008).

\bibitem{gau2} E. Kluk, M. F. Herman, and H. L. Davis, J. Chem. Phys. {\bf 84},
326 (1986).

\bibitem{gau3} W. T. Duncan and T. N. Truong, J. Chem. Phys. {\bf 103},
9642 (1995).

\bibitem{novaes1} M. Novaes, J. Math. Phys. {\bf 46},
102102 (2005).

\bibitem{novaes2} M. Novaes and M. A. M. de Aguiar, Phys. Rev. A {\bf 72},
032105 (2005).


\end{thebibliography}
\end{document}